\documentclass{JHEP3}

\usepackage{latexsym}

\usepackage{amsfonts}
\usepackage{amsmath}

\usepackage{indentfirst}

\def\a{\alpha} 
                                  
\def\d{\delta}
                                  
\newcommand{\ep}{\epsilon}
\newcommand{\g}{\gamma}                         
\newcommand{\G}{\Gamma}
\def\k{\kappa}
\def\la{\lambda}
\def\La{\Lambda}
\def\om{\omega}

\def\v{\upsilon}

\def\cB{{\cal B}}
\def\cC{{\cal C}}
\def\cD{{\cal D}}  

\def\cH{{\cal H}}
\def\cK{{\cal K}}
\def\cR{{\cal R}}
\def\cU{{\cal U}}
\def\cO{{\cal O}}

\def\cS{{\cal S}}

\def\R{\mathbf{R}} 
\def\M{\mathbf{M}}

\def\me{\mathbf{e}}                                

\def\bN{{\mathbb{N}}}
\def\bZ{{\mathbb{Z}}}
\def\bR{{\mathbb{R}}}
\def\S{\mathbb{S}}
\def\1{\mathbb{I}}

\def\beq{\begin{equation}} 
\def\eeq{\end{equation} }
\newcommand{\beqa}{\begin{eqnarray}}
\newcommand{\eeqa}{\end{eqnarray}}
\newcommand{\barr}{\begin{array}}
\newcommand{\earr}{\end{array}}
\newcommand{\bit}{\begin{itemize} }
\newcommand{\bgt}{\begin{gather} }
\newcommand{\egt}{\end{gather}}
\newcommand{\bal}{\begin{align}}
\newcommand{\eal}{\end{align}}
\newcommand{\lb}{ \left( } 
\newcommand{\rb}{ \right) }
\newcommand{\lan}{ \langle }
\newcommand{\ran}{ \rangle }

\newcommand{\aut}{automorphism  } 
\newcommand{\autn}{automorphism }
\newcommand{\auts}{automorphisms  }
\newcommand{\autsn}{automorphisms}
\newcommand{\rep}{representation }
\newcommand{\repn}{representation}
\newcommand{\reps}{representations }
\newcommand{\repsn}{representations}
\newcommand{\irrep}{irreducible representation }

\newcommand{\irreps}{irreducible representations }
\newcommand{\irrepsn}{irreducible representations}
\newcommand{\hws}{highest weight state }

\newcommand{\hw}{highest weight }

\newcommand{\wrt}{with respect to }

\newcommand{\kp}[1]{k_{\ep_{#1}}} 
\newcommand{\km}[1]{k_{- \ep_{#1}}}
\newcommand{\gt}[1]{\mathfrak{#1}}
\newcommand{\ovl}[1]{\overline{#1}}
\newcommand{\tx}[1]{\textrm{#1}}
\newcommand{\ciut}[1]{\tiny$#1$}
\newcommand{\nl}{\newline} 

\newcommand{\non}{\nonumber\\}

\def\det{{\rm det}}
\def\tr{{\rm tr}}

\def\faffN{P^{\k}_{+}(A_N)}
\def\qAN{\cU_q (A_N)}
\def\hAN{\cU_h (A_N)}
\def\qANex{\cU^{\rm{ext}}_q (A_N)}
\def\reAN{\textrm{REA}_q(A_N)}
\def\qU2{\cU_q (\mathfrak{su}(2))}
\def\resu2{\textrm{REA}_q(\mathfrak{su}(2))}
\def\must{\overset{!}{=}}
\def\x{\times}
\def\ox{\otimes}
\def\lx{\ltimes}

\def\spun{\textrm{span}}

\def\Mat{\textrm{Mat}}
\def\id{\textrm{id}}

\def\Rep{\mathcal{R}ep}
\def\Out{{\cal O}ut}
\def\sOut{s{\cal O}ut}
\def\Aut{{\cal A}ut}
\def\Int{{\cal I}nt}
\def\act{\vartriangleright}
\def\ggt{\mathfrak{g}}
\def\faff{P^{\k}_{+}(\ggt)}

\def\kmg{\hat{\mathfrak{g}}_{\k}}

\def\Tr{\tx{Tr}}

\def\dziel{\ \vert \ }

\def\too{\longrightarrow}

\title{Quantum Matrix Models for Simple Current Orbifolds~{}\footnote{Work
supported
by Polish State Committee for Scientific Research (KBN) under contract
2 P03B 001 25 (2003-2005)}}

\author{Jacek Pawe\l czyk and Rafa\l ~R. Suszek \\ Institute of
Theoretical Physics \\ Warsaw University \\ Ho\.za 69, PL-00-681
Warsaw \\ E-mail: \email{pawelc@fuw.edu.pl, suszek@fuw.edu.pl}}

\abstract{An algebraic formulation of the stringy geometry on simple current
orbifolds of the WZW models of type $A_N$ is developed within the framework of
Reflection Equation Algebras, $\reAN$. It is demonstrated that $\reAN$ has the 
same set of outer automorphisms as the corresponding current algebra
$A_N^{(1)}$ which is crucial for the orbifold construction. The CFT monodromy 
charge is naturally identified within the algebraic framework. The ensuing
orbifold matrix models are shown to yield results on brane tensions and 
the algebra of functions in agreement with the exact BCFT data.}

\keywords{WZW models, simple current orbifolds, quantum groups, reflection
equation algebras}

\preprint{XXX}

\begin{document}

\newpage

\section{Introduction.}

Physics of D-branes has long been a subject of intense study\footnote{See 
\cite{schom} for a detailed review and an exhaustive list of references.},
driven by the motivation to obtain new insights into the structure of the
moduli space of string theory proper and to better understand the emergence
of an essentially stringy geometry and gauge dynamics within any
field-theoretic or matrix model approach to the propagation of strings in
curved gravitational backgrounds with fluxes. An example of such a background 
is a compact Lie group $G$ (\cite{wzwitt} and \cite{gep-witt}), or a quotient 
thereof, known to support a nontrivial Kalb--Ramond field in a conformally 
invariant theory. Maximally symmetric (or untwisted) D-branes on $G$ have been 
shown to localise stably (\cite{flux-stab}) around a discrete set of conjugacy 
classes (\cite{geowzw}) and are enumerated by dominant integral affine weights 
$\La$ from the fundamental affine alcove $\faff$\footnote{Here, as anywhere else 
in this paper, we restrict our analysis of the pre-orbifolding CFT to the 
diagonal case in which Cardy's classification of boundary states applies 
(\cite{bcft-pion}).}, i.e. we have:
\beq
\tx{(untwisted) D-branes on $SU(N+1)$} \quad \sim \quad
\Rep_{\substack{integrable  
\\ highest \ weight}} \lb A^{(1)}_N \rb := \bigoplus_{\La \in \faffN} R_{\La},
\end{equation}
in the case of interest, with
\beq
\faffN := \left\{ \La = \sum_{i = 1}^N \la_i \La^i \in P^*(A_N) \quad
  \bigg\vert  
\quad \forall_{i \in \ovl{1,N}} \ : \ \la_i \in \bN \ \ \land \ \ \sum_{i = 
1}^N \la_i \leq \k \right\},
\end{equation}
where $P^*(A_N)$ is the weight space of $A_N$ and $\La^i$ are the fundamental
weights. The above defines the particular class of WZW manifolds whose
orbifolds shall be the subject matter of the present paper.

In \cite{ps} a (quantum) matrix model of branes in WZW models was constructed,
based on certain quantum algebras called Reflection Equation (RE) algebras.
The latter have representation theory closely related to that of $U_q(A_N)$,
which  
- in turn - is known to have irreducible \hw \reps such as those of
$A_N^{(1)}$.  

In this paper we extend the former construction  to a
new class of  WZW backgrounds, namely - the so-called simple current
 orbifolds of the WZW geometries of type $A_N$. Simple currents are known to
 form an Abelian group, contained as a proper subgroup in the group of outer
 automorphisms of $A_N^{(1)}$ (we shall denote the latter by
 $\Out \lb A_N^{(1)} \rb$). Thus an orbifolding of the corresponding matrix model
 based on $\reAN$ requires that the algebra has the same group of outer
 automorphisms as the affine algebra. 
Indeed, we show that $\Out\lb \reAN \rb \cong \Out \lb A_N^{(1)} \rb$ and 
moreover, that the action of both sets of \auts on branes is identical. As a
result, $\Out \lb \reAN \rb$ acquires a geometrical  
meaning necessary to perform the construction of the orbifold. Furthermore, 
it is straightforward to identify the CFT monodromy charge in $\reAN$ so that the 
orbifolding itself admits a purely algebraic and hence natural realisation in the 
framework developed, reminiscent of the original approach to orbifold models 
advanced in \cite{dix}.
 
At this stage we may already outline and assess the content of the present
paper. In Sect.\ref{rea}. we briefly describe RE algebras, their \auts and their
relation to branes, shifting some technicalities to the appendices. In the following 
section we apply our results in a construction of quantum orbifolds and discuss their 
properties in some detail, whereby we also arrive at a particularly straightforward
interpretation of the monodromy projection (known from BCFT) accompanying simple 
current orbifolding. These two sections form the core of the paper. In 
Sect.\ref{example} we present an explicit example of branes on the $SU(2)/\bZ_2$ 
orbifold.

The all-important details and notation can be found in the Appendices. The
first one, App.\ref{app:qBCFT}, introduces the necessary BCFT background. In 
App.\ref{sec:deform} we present a path leading from BCFT to the quantum matrix model. 
The remaining Appendices contain some relevant information on the algebras: 
$\qANex$ and $\reAN$ and their representations.

\section{The Reflection Equation Algebras and branes.}\label{rea}

 Similarities in the representation theory of the algebras: 
$A_N^{(1)}$ and $\cU_q(A_N)$ ($q$ being a root of unity) are widely 
appreciated. There are, however, important differences as well. One of 
them is the outer \auts group\footnote{$\Aut$ is the set of all \autsn, 
with the identity adjoined to endow it with the group structure. $\Int$ 
is the set of inner \autsn.}, $\Out = \Aut/\Int$. For $A_N^{(1)}$ the 
latter reflects the symmetries of the appropriate Dynkin diagram, $\Out 
\lb A_N^{(1)} \rb = \bZ_{N+1} \lx \bZ_2$, while for $\qAN$ we have 
$\Out(\cU_q(A_N))=\bZ_2^N \lx \bZ_2$. The difference is crucial in the 
context of simple current orbifolding since - according to \cite{fux} - 
the group generated by simple currents under OPE is precisely the 
strictly affine factor of $\Out \lb A_N^{(1)} \rb$, that is - $\bZ_{N+1}$.

In this section, we investigate the so-called Reflection Equation Algebra 
(\cite{skl,kss}), closely related to a modification of $\cU_q(A_N)$ named 
the extended quantum universal enveloping algebra and denoted as $\qANex$. 
As we explore its \rep theory it shall become clear that, beside 
representations, $\reAN$ shares with $A_N^{(1)}$ the set of outer 
automorphisms $\bZ_{N+1} \lx \bZ_2$, with - as indicated by our results - 
the same geometrical meaning of the $\bZ_{N+1}$ factor as in the affine 
setup. The last property is crucial for constructing orbifold models based 
on $\reAN$ in strict analogy with the WZW orbifolds discussed in 
App.\ref{app:qBCFT}.

Recall that the Reflection Equation Algebra, $\reAN$, is the algebra 
generated by the operator entries of the matrix $\M$ determined by the
celebrated Reflection Equation (\cite{skl}):
\beq\label{REprim}
\R_{12} \M_1 \R_{21} \M_2 = \M_2 \R_{12} \M_1 \R_{21}.
\end{equation}
The independent central terms of this algebra are given by 
(\cite{arnba}) 
\bgt
\mathfrak{c}_k := \tr_q ( \M^k ),\qquad k \in \ovl{1,N}, \label{re-ce} 
\\ \non
\cK := M_{N+1,N+1}^{\k_N} = e^{-\frac{2 \pi i}{N+1} \sum_{n = 
1}^N n H_n}, \label{re-K}
\end{gather}
where on the r.h.s. of the last formula we used a specific representation 
of $M_{N+1,N+1}$, to be justified presently. We interpret $M_{ij}$'s as 
coordinate ``functions'' on a quantum manifold, thus $\mathfrak{c}_k$ 
represent algebraic constraints on positions of submanifolds to be 
associated with branes. Amazingly, the last Casimir, $\cK$, is related to 
the BCFT monodromy charge as one can easily check by comparing the above with 
\eqref{monod}. 

There is one more scalar, the quantum determinant, to be set to some 
specific value:
\beq\label{qdet-scal}
\det{}_q \M \sim \1.
\end{equation}

As an immediate consequence of \eqref{REprim} and \eqref{qdet-scal}
we get $\M \too e^{\frac{2 \pi i L}{N+1}} \M, \ L \in \bZ_{N+1} \setminus 
\{0\}$ as outer automorphisms of $\reAN$. Here we shall be interested in finite 
dimensional \reps of $\reAN$ induced through the homomorphisms: $\reAN 
\hookrightarrow \qANex \hookrightarrow \hAN$ from highest weight \irreps of the 
latter algebra, with a non-vanishing quantum dimension (see App. \ref{sec:rea}).
We shall demonstrate that the above homomorphisms give rise to $\Out(\reAN) 
= \bZ_{N+1} \lx \bZ_2$, with the $\bZ_{N+1}$ factor realised as above and 
the remaining $\bZ_2$ being the standard mirror symmetry of the Dynkin 
diagram of $A_N$.

The action of the outer \auts \eqref{phex}-\eqref{conphex} (with the 
generator denoted by $\eta$) reads
\beq\label{out-m}
\eta: M_{ij} \to e^{\frac{2 \pi i}{N+1}} M_{i j},
\eeq
which implies
\beq\label{out-ck}
\eta \ : \ \mathfrak{c}_k  \to e^{ \frac{2 \pi i k}{N+1}}
\mathfrak{c}_k \quad , \quad \cK \to e^{ \frac{2 \pi i \k }{N+1}}
\cK.
\eeq
It appears that the above is the same as the left action of the inverse of the 
Casimir \eqref{re-K},
\beq\label{KLM}
\cK^{-1} \act_L M_{ij} := \pi_{ik} \lb \cK^{-1} \rb M_{kj} = e^{\frac{2 \pi 
i}{N+1}} M_{ij} \equiv \eta \lb M_{ij} \rb.                                         
\end{equation} 
This is just an algebraic analogue of the ordinary geometrical left regular 
action: 
\beq\label{ccact}
g \too \g g, \qquad g \in SU(N+1) \ , \ \g \in Z(SU(N+1)) \cong \bZ_{N+1}.
\end{equation}

Next, we turn to $\Out ( A_N^{(1)})$. To begin with, recall that $\bZ_{N+1} 
\subset \Out \lb A_N^{(1)} \rb$ can be identified with a (discrete rotational) 
symmetry of the fundamental affine alcove $\faffN$ and also with a cyclic 
symmetry of the extended Dynkin diagram of $A^{(1)}_N$. Moreover, $\bZ_{N+1}$ 
has a deep geometrical meaning: it is isomorphic to the centre of the group
$SU(N+1)$, thus orbifolding by $\bZ_{N+1}$ means taking the quotient 
$SU(N+1)/\bZ_{N+1}$ \wrt the natural action \eqref{ccact}. In the Dynkin basis, 
the action of its generator on $\faffN$ takes the following form:
\beq\label{affrot}
\om^{*}_N \ : \ \left[ \la_1 , \la_2 , \ldots , \la_N \right] \to
\left[ \la_2 , \la_3 , \ldots , \la_N , \k - \sum_{i = 1} \la_i
\right].
\end{equation}
It results in the rescaling of the eigenvalues of the scalars             
\eqref{re-ce}-\eqref{re-K}:           
\beq\label{aff-rot-sc}
\gt{c}_k \to e^{\frac{2 \pi i k}{N+1}} \gt{c}_k \quad , \quad
k_{\ep_{N+1}}^{2\k_N} \to e^{\frac{2 \pi i \k}{N+1}}
k_{\ep_{N+1}}^{2\k_N},
\end{equation}
which matches \eqref{out-ck} exactly. Thus the action of 
$\bZ_{N+1} \subset \Out \lb A_N^{(1)} \rb$ on $\gt{c}_k$ is the same as that 
of $\bZ_{N+1} \subset \Out \lb \reAN \rb$. 

In order to directly relate both sets of automorphisms in a physically 
meaningful manner we need to modify our former definition of the brane in the 
algebraic setup. In \cite{ps}, string theory branes in the $SU(N+1)$ group 
manifold were associated with \irreps of $\reAN$. Here we propose to take the 
specific $(N+1)$-fold direct sum of \irrepsn:
\beqa
\cB_{\La} := \bigoplus_{L = 0}^N R^L_{\lb \om^{*}_N \rb^L \La}
\label{brane}
\end{eqnarray}
as describing the brane assigned to the weight $\La$. The 
(quantum) manifold of $SU(N+1)$ is thus ''foliated'' by the set of branes 
corresponding to all $\La\in \faffN$. Notice that all the
summands in \eqref{brane} have the same Casimir  
eigenvalues which is necessary for the correct geometrical picture.
The advantage of 
\eqref{brane} is that\footnote{$\eta^* R^L_\La = R^{L+1}_\La$ where $R^{0}_\La 
\equiv R^{N+1}_\La$ and where we have introduced the action $\eta^*$ on weights 
induced by \eqref{out-m} in an obvious manner.}              
\beq\label{eta-brane}
\eta^* \cB_{\La} = \cB_{\om_N^* \La},                                      
\eeq   
i.e. the corresponding branes are isomorphic (the same Casimir eigenvalues 
and the same algebra of functions). In this setting $\eta^*$ is indistinguishable 
from $\om_N^*$ and so it acquires an analogous geometrical meaning, which
shall be  
crucial in our subsequent considerations. In particular, it allows us to
define the  
orbifold by dividing out the action of $\bZ_{N+1}$ generated by
$\eta$. Orbifolding  
will ``glue'' branes with $\om_N^*$-conjugate Casimirs.

For some non-generic weights (``fixed point'' weights) we may have 
$(\om^*_N)^D \La_{FP} = \La_{FP}$ (with $\frac{N+1}D =: n \in \bN$), i.e. 
$(\eta^*)^D \cB_{\La_{FP}} = \cB_{\La_{FP}}$. When evaluated on those 
``fixed point'' weights $\La_{FP}$, some of the Casimirs vanish. This will 
lead to certain new interesting phenomena discussed in Sec.\ref{sub:fpr}.

As an aside, we note here that as we introduce \eqref{brane} the requirement 
of consistency with the BCFT data for the pre-orbifolding WZW models 
necessitates a reformulation of the original quantum matrix model, which we 
now take in the general form\footnote{Cp \cite{ps}.}:
\bgt
S^{eff}_{q,A_N} := \frac{T_0}{N+1} \Tr_q \left[ \1 + \tx{cov}_{\qAN} ( \M ) 
\right]. \label{nactAN}
\end{gather}
Here $\tx{cov}_{\qAN} ( \M )$ denotes an unspecified function of the matrix 
variable $\M$, covariant under the action of $\qAN$ described in \cite{ps}. 
The action \eqref{nactAN} is to be evaluated on representations of the kind 
\eqref{brane}.

Equipped with these result, one can now try to perform constructions of 
branes in simple current WZW orbifold models, e.g. on the manifold 
$SU(N+1)/\bZ_{N+1}$. 

\section{The quantum orbifold.}\label{q-orb-mm}

The present section is central to our paper. It contains an explicit 
construction of the orbifold. We also demonstrate the appearance of new 
quantum geometries. The latter describe fixed point D-branes of the simple 
current $\bZ_{N+1}$-orbifold of the WZW model of type $A_N$.

In orbifolding, the Casimir $\cK$ proves to play a prominent r\^ole. It was 
used in \eqref{KLM} to implement the appropriate action of $Z_{N+1}$ on the 
coordinate variables $M_{ij}$. Upon restricting to the set of corresponding 
$\cK$-invariant monomials within $\reAN$ we recover the orbifold algebra in 
the form\footnote{The discussion below is confined to the maximal 
$\bZ_{N+1}$-orbifold for the sake of concreteness only but the conclusions 
drawn are of general validity.}:
\beq\label{orb-gen}
\reAN / \bZ_{N+1} = \spun \left\langle M_{i_1 j_1} M_{i_2 j_2} \cdots
M_{i_{N+1} j_{N+1}} \right\rangle/I_{RE,\det_q},
\end{equation}
where $I_{RE,\det_q}$ is the ideal generated by \eqref{REprim} and 
\eqref{qdet-scal}, rephrased in terms of the invariant monomials. Clearly, 
the orbifold algebra is unital.

Determining the geometry of D-branes in the quantum-algebraic setup requires 
two pieces of data: the algebra of ``functions'' and its \rep theory. Thus 
it is legitimate to consider - in addition to \eqref{KLM} - the action of 
$\cK$ on \eqref{brane}:
\beq\label{K-mon}
\cK \act v_{\La} = R_{\La} (\cK) v_{\La} = e^{\frac{2 \pi i}{N+1} \cC(\La)} 
v_{\La}, \qquad v_{\La} \in \cB_{\La}.
\end{equation}
In this case, projecting onto the subset of $\cK$-invariant \reps is 
tantamount to imposing the condition: 
\beq\label{conj0}
\cC(\La)=0 \mod N+1. 
\end{equation}
The latter means that the monodromy charge \eqref{mono-exp} of the \rep 
considered should vanish, in accordance with the BCFT results (cp 
App.\ref{app:qBCFT}). 

Altogether, the Casimir $\cK$ is seen to realise the action of $\bZ_{N+1} 
\subset \Out \lb \reAN \rb$ both at the level of the algebra and that of 
the associated \rep theory, sending us from the original geometry - as 
given by $\reAN$ and $\Rep_{\tx{ind.}} ( \reAN )$ - to the orbifold one 
upon dividing out its left action on the two components.

In order to be able to assess the structure encoded in \eqref{orb-gen}
we distinguish within the set of $\bZ_{N+1}$-invariants the following 
subsets:
\beqa
\{ M^{N+1} \} \ , \ \{ \gt{c}_1 M^{N}\}, \dots \ \{ \gt{c}_N M_{i j} \ , 
\ \gt{c}_{N-1} \gt{c}_1 M_{i j} \ , \ \gt{c}_{N-2} \gt{c}_2 M_{i j} \ , \
\ldots \ , \ \gt{c}_1^N M_{i j}\} \ , \ \{ \det_q \M \sim \1 \}, \non
\label{deCas}
\eeqa
where $M^p$ symbolically denotes an arbitrary monomial in $M_{ij}'s$ of 
degree $p$ as in \eqref{orb-gen}. The above decomposition shows that the 
orbifold geometry associated to a generic weight (on which some $\gt{c}_k 
\neq 0$) is as in the pre-orbifolding case i.e. it is generated by 
$M_{i j}$'s (modulo $I_{RE,\det_q}$, as usual). For these weights, we may 
take the definition of the brane on the orbifold to be as in \eqref{brane}, 
supplemented by the constraint \eqref{conj0}. The fixed point geometries, 
on the other hand, display an altogether different structure, analysed at 
some length in the next section.

It ought to be remarked at this point that the orbifolding described cuts 
the range of admissible affine weight labels $\La$ to $\faffN/\bZ_{N+1}$, 
with the division determined by the action of $\om^*_N$ (equivalent to
$\eta^*$ in this case) together with the monodromy projection \eqref{conj0}.
The resulting set spans an $N$-dimensional solid within $\faffN$ with 
vertices defined by the vectors: 
\beq\label{elem-D}
\mathcal{D} \ : \ \left\langle 0 , \frac{\k}{2} \La^i , \frac{\k}{3} \lb 
\La^j + \La^{j+1} \rb , \frac{\k}{3} \lb \La^1 + \La^N \rb , \frac{\k}{N+1} 
\sum_{k = 1}^N \La^k \right\rangle^{i \in \ovl{1,N}, \ j \in \ovl{1,N-1}}. 
\end{equation}    
One should note, in particular, the presence of the central weight
$\tfrac{\k}{N+1} [1,1,\ldots,1]$ (Dynkin label notation) in \eqref{elem-D}. 
It is an actual element of $\faffN$ whenever $N+1 \dziel \k$ ($N+1$ divides 
$\k$) and an actual element of $\faffN/\bZ_{N+1}$ iff $N+1 \dziel \tfrac{N 
\k}{2}$.

\subsection{New geometries in $\reAN / \bZ_{N+1}$.}\label{sub:q-orb-alg}

 The non-generic features of orbifold geometries come to the fore 
whenever there is a fixed point of the action of $(\om_N^{*})^D$ on 
$\faffN$ for some $1 \leq D < N+1$. In this case some of $\gt{c}_k$ 
vanish\footnote{Recall that we have imposed $\cK \must 1$.}, as follows 
from \eqref{aff-rot-sc}: since the $N+1$ independent Casimir eigenvalues 
pick up the $N+1$ independent phase factors under the simple current rotation 
$\om_N^{*}$ the corresponding weight $\La_{FP}$ will be stabilised by 
$(\om_N^{*})^D$ iff the Casimirs whose eigenvalues do change under the latter 
vanish. The same conclusion holds for $\eta$ due to \eqref{eta-brane}. Among 
the nontrivially stabilised weights there migth be a distinguished one, 
sitting in the geometric centre of the $N$-simplex of the fundamental affine 
alcove (cp \eqref{elem-D} and subsequent remarks).  

We want to gain some insight into the structure of fixed point algebras. To 
these ends consider a restriction of the algebra of $\bZ_{N+1}$-invariants to 
the \rep \eqref{brane} of $\reAN$ associated with a given weight $\La$ and 
characterised by the following set of non-zero Casimir eigenvalues:
\beq\label{Cas-n0}
C_{k_1} , C_{k_2} , \ldots , C_{k_K} \neq 0 
\end{equation}
where $C_j := \gt{c}_j\vert_{\cB_{\La}}, \ j \in \ovl{1,N}$ 
and $C_i = 0$ for all $i \notin \{ k_1 , k_2 , \ldots , k_K \}$.            

{\bf We then claim that the corresponding coordinate subalgebra is generated 
by independent monomials of degree $n$ such that 
\beq\label{div-N-alg}
n = \gcd \lb N + 1 , k_j \rb_{j \in \ovl{1,K}}.
\end{equation}}
Here is the proof of our claim. Let $n$ be as in \eqref{div-N-alg}.
Given \eqref{Cas-n0}, one can form a polynomial of degree $N+1$ 
(a $\bZ_{N+1}$-invariant) $\prod_{j=1}^K \gt{c}_{k_j}^{p_j} M^p$, where $M^p$ 
is an arbitrary monomial of degree $p$ and $n|p$. Thus one can effectively use 
the monomials $M^p$ to generate the orbifold algebra. Next one takes products 
$(M^p)^l$ with $l > 0$ the smallest number such that $(M^p)^l = M^{p_1} \det_q 
\M$ for some non-zero $p_1$. Then $p_1 < p$ and $p_1$ is divisible by $n$.
Continuing the procedure one concludes, after a finite number $S$ of steps, 
that it is monomials $M^{p_S}$ that generate the algebra, with $p_S \dziel N+1$. 
One can subsequently perform analogous reduction \wrt each of the non-vanishing 
Casimirs, whereby one finally reaches the desired conclusion \eqref{div-N-alg}.

The monomials $M^n$ are invariant under $\bZ_n \subset \bZ_{N+1}$ generated 
by $\eta^D$. The subgroup $\bZ_n$ is the stabiliser of $\La$ within 
$\bZ_{N+1}$. In the distinguished case of the central weight (and 
exclusively in that case) all $\gt{c}_k=0$ and we recover the full $\bZ_{N+1}$
as the stabiliser, with the algebra of monomials of degree $N + 1$ as the 
generating one. We expect the emergence of essentially new quantum geometries 
whenever the number \eqref{div-N-alg} is greater than one. These quotient 
geometries are described by proper subalgebras of $\reAN$ and - as argued below 
- have a reduced \rep theory. We plan to dwell on this subject in a separate 
publication.

\subsection{The fixed point resolution.}\label{sub:fpr}

Here we shall deal with one of the ``fixed point'' weights $\La=\La_{FP}$ which
defines the group $Z_n$ as above. Following the definition \eqref{brane} we
consider
\beqa
\bigoplus_{L = 0}^N R^L_{ (\om^{*}_N)^L \La}=\bigoplus_{x = 0}^{n-1}
\bigoplus_{L = 0}^{D-1} R^{L + xD}_{ (\om^{*}_N)^L \La}
\label{brane-try}
\end{eqnarray}
where the equality holds due to $(\om^*)^D \La = \La$ ($nD=N+1$). The space of 
functions on the ``fixed point'' brane is generated by the monomials $M^n$. 
Recalling that $R^{L + lD}_\La=R^0_\La\otimes e^{\frac{2 \pi i L}{N+1}+ 
\frac{2 \pi i l}{n}}$ (see App.\ref{sub:rep-out-rea} and App.\ref{app-inequiv}) 
we thus obtain, symbolically, $\lb R^{L + lD}_\La \rb^n = \lb R^{L}_\La \rb^n$. 
From this point of view the $x$-dependence in \eqref{brane-try} drops out 
completely on passing to the orbifold geometry. Consequently, we could choose
\beq
b_\La = \bigoplus_{L = 0}^{D-1}  R^{L}_{ (\om^{*}_N)^L \La}
\eeq
as a possible definition of the (fractional) brane. The latter satisfies
$\lb \eta^* b_\La \rb^n = \lb b_{\om_N^*\La} \rb^n$ (see \eqref{eta-brane}), 
which is necessary for the geometrical meaning of $\eta$.

On the other hand, the BCFT results indicate that for each such a fixed point
weight there should be $n$ different branes. This means that we need to 
distinguish different $x$ labels somehow. We shall do it by introducing a 
certain cross product extension of the algebra generated by $M^n$. Given the 
action of the $Z_n$-generator 
\beq
(\eta^*)^{D} \ \lb \bigoplus_{L = 0}^{D-1} R^{L + xD}_{ (\om^{*}_N)^L \La} 
\rb =\bigoplus_{L = 0}^{D-1} R^{L + (x+1)D}_{ (\om^{*}_N)^L \La},
\eeq
we see that the original module \eqref{brane-try} splits into 
$(\eta^*)^{D}$-eigenspaces as
\beq\label{split}
\bigoplus_{L = 0}^N R^L_{ (\om^{*}_N)^L \La} =  
\bigoplus_{y = 0}^{n-1} P_y \lb \bigoplus_{L = 0}^{D-1} \bigoplus_{x =
  0}^{n-1} R^{L + xD}_{(\om^{*}_N)^L \La} \rb,
\eeq
where the projectors $P_y$ onto respective eigenspaces are
\beq\label{proj}
P_y := \frac{1}{n} \sum_{m=0}^{n-1} e^{- \frac{2 \pi i y m}{n}} (\eta^*)^{D 
\cdot m}, \qquad P_{y}P_{z} = \d_{y,z} P_{y},\quad \ y,z \in \ovl{0,n-1}.
\eeq
We can then identify, for arbitrary $y \in \ovl{0,n-1}$, the 
$\cB_\La^y $--brane:
\beq\label{bres}
\cB_\La^y := P_y \lb \bigoplus_{L = 0}^{D-1}\bigoplus_{x = 0}^{n-1} R^{L +
  xD}_{(\om^{*}_N)^L \La} \rb.
\eeq
The crucial property of $\cB_\La^y$ is $(\eta^*) \cB_\La^y = \cB_{\om_N^*\La}^y$ 
from which $(\eta^*)^{D} \cB_\La^y = \cB_{\La}^y$ follows. 
We may equivalently rewrite \eqref{bres} as
\beq
\cB_\La^y = b_\La \ox p_y,
\eeq
with $p_y$ - the $y$--th irreducible, one-dimensional \rep of $\bZ_n$. 
Consequently, we define the space of functions on the $y$--th brane 
($y=0,1,...,n-1$) to be generated by
\beq\label{cp-alg}
M^n \otimes P_y (\cK),
\eeq
where $P_y(\cK)$ is as in \eqref{proj} but with $\eta^*$ replaced by $\cK$, 
and $p_y \lb \cK^D \rb = e^{\frac{2 \pi i y}{n}}$.                                        
The crossed product structure is completed by defining the product of the 
generators \eqref{cp-alg}. The latter is inherited from the 
following\footnote{Cp \cite{maj}.} crossed product $(M_{ij} \otimes \cK^{D \cdot 
m}) \cdot (M_{i'j'} \otimes \cK^{D \cdot m'})= M_{ij} (\cK^{D \cdot m} \act_L 
M_{i'j'}) \otimes \cK^{D \cdot (m+m')}$ and reads (symbolically) 
\beq
(M_1^n\otimes P_x(\cK))\cdot(M_2^{n} \otimes P_y(\cK))= \d_{x,y} \ M_1^n M_2^{n} 
\otimes P_x(\cK).
\eeq
The algebra thus defined, together with its \rep theory, describes the so-called 
fractional orbifold branes, discussed in the present context e.g. in 
\cite{matsu}. Their existence was inferred, in particular, from the structure of 
the fixed point boundary OPE algebra, encoding the crossed product extension of 
the relevant algebra of functions, as argued in \cite{matsu}. Finally, let us add 
that the emergence of the crossed product extension of the algebra of functions 
at fixed points of the orbifold action is a general feature of matrix models of 
orbifold geometries, reflecting the existence of fractional branes (\cite{orb-gen}).

\subsection{Orbifold brane tensions.}\label{sub:q-orb-ten}

An important and nontrivial test of the construction presented above
is the computation of tensions of the orbifold branes within the
framework developed. In so doing we follow the scheme advanced in the
original papers, \cite{ps}, that is - we compute the tension of a
brane labelled by the weight $[\La] \in \cD \subset \faffN$ by evaluating 
the first term of the effective action \eqref{nactAN} on the associated
\repn, $\cB_{[\La_{FP}]}^y$, for a fixed point weight $\La = \La_{FP}$, or 
$\cB_{\La} =: \cB^{off}_{[\La]}$ for an off-fixed-point one (clearly, the 
orbifolding of off-fixed-point modules produces modules isomorphic to the 
original (pre-orbifolding) ones). Beside giving rise to the aforementioned 
crossed product extension of the fixed point algebra of $\bZ_{N+1}$-invariants, 
the latter yields the correct result for the tension of resolved fixed point 
branes $\cB_{[\La_{FP}]}^y$:
\beq\label{orb-ten-fp}
\frac{T_0}{N+1} \Tr_q\big\vert_{\cB_{[\La_{FP}]}^y} \1 = \frac{1}{N+1} 
\cdot D \cdot \Tr_q\big\vert_{R^0_{\La_{FP}}} \1 = \frac{1}{n} 
\mathcal{E}_{\La_{FP}}, \qquad y \in \bZ_n, \quad n = \vert \cS(\La_{FP}) 
\vert,
\end{equation}
falling in perfect agreement with the BCFT data (cp 
\eqref{frac-ten-bcft}). It is supplemented by an equally satisfactory 
result for generic (off-fixed-point) branes $\cB^{off}_{[\La]}$:
\beq\label{orb-ten-nfp}
\frac{T_0}{N+1} \Tr_q\big\vert_{\cB^{off}_{[\La]}} \1 = \frac{T_0}{N+1} 
\cdot (N+1) \cdot \Tr_q\big\vert_{R^0_{\La}} \1 = \mathcal{E}_{\La},
\end{equation}
an immediate consequence of the lack of an internal mechanism of
spectrum reduction in this last case.

\section{An example: branes on $\bR P^3_q$.}\label{example}

Below we detail the particularly simple example of the antipodal
$\bZ_2$-orbifold of the quantum matrix model for $\gt{su}_{\k}(2), \
\k \in 4 \bN^*$. The model is exemplary in that it develops precisely 
along the lines discussed in the previous sections, hence we restrict 
ourselves here to its q-geometric interpretation, relating our results 
to some well-established mathematical constructs of \cite{pod} and 
\cite{rp2}. The BCFT properties of the corresponding branes were 
discussed in \cite{matsu,cou}.

\subsection{The setup.}

Our starting point shall be the RE algebra suggested in \cite{ps} as giving 
a plausible compact description of the quantum D-brane geometry on WZW 
group manifolds. Since we aim at describing the stringy geometry of 
the antipodal orbifold of $SU(2)$ we will focus on the RE algebra generated by 
the operator entries of a matrix $\M$ subject to the reflection equation 
\eqref{REprim} in which $\R$ is the standard (Cp \cite{maj}) universal 
$\cR$-matrix of $\qU2$ in the bi-fundamental \repn,                                               
\beq\label{rmat}
\R = \sum_{i,j,k,l = 1}^2 R^{ij}_{\; kl} \me_{i k} \ox \me_{jl} = q 
\me_{1 1} \ox \me_{1 1} + q \me_{2 2} \ox \me_{2 2} + \me_{1 1} \ox 
\me_{2 2} + \me_{2 2} \ox \me_{1 1} + \la \me_{1 2} \ox \me_{21}. 
\end{equation}
Having parameterised $\M$ as
\beq\label{q-Pauli}
\M = \left( \begin{array}{cc} M_4 - i q^{-2} M_0 & i q^{- \frac{1}{2}}
\sqrt{[2]_q} M_{-1} \\
- i q^{- \frac{3}{2}} \sqrt{[2]_q} M_1 & M_4 + i M_0
\end{array} \right), 
\end{equation}
we may write down the additional Casimir constraint:
\beq \label{req}
r^2 \1 \equiv \det{}_q \M = M_4^2 + M_0^2 - q^{-1} M_1
M_{-1} - q M_{-1} M_1,
\end{equation}
in a natural way, with $r^2$ interpreted as the radius squared of the 
group manifold, $\S_q^3$, which we set roughly proportional to the 
level of the underlying WZW model (cf \cite{ps}),
\beq
r \approx \sqrt{\a' \k_1}.                                                       
\end{equation}
The RE \eqref{REprim} now takes a manageable (component) form ($i \in 
\{ 0, 1, -1 \}$):
\beqa\label{recom}
[ M_4 , M_i ] = 0 \quad &,& \quad 
M_0 M_{-1} = q M_{-1} ( q M_0 - i \la M_4 ), \non \non
M_{1} M_0 = q ( q M_0 - i \la M_4 ) M_{1} \quad &,& \quad 
M_{1} M_{-1} - M_{-1} M_{1} = \la M_0 ( M_0 - i M_4 ).
\eeqa
The above algebra, further constrained by \eqref{req}, is easily seen to
reproduce (after diagonalising the central element $M_4$ and some 
trivial rescalings) the celebrated Podle\'s' spheres, $\S^2_{q,c}$, with 
the parameter $c$ essentially determined by the value of $M_4$ (cf 
\cite{pod}). 

Bearing in mind the clear geometric picture of the REA, we should
expect that the general q-\aut \eqref{out-m} corresponds to the standard
antipodal identification of "points" on $\S_q^3$. In order to verify that 
we consider the embedding \eqref{re-to-u} and compute
\beq\label{MinUq}
\cK^L \act \M = r e^{- \pi i L} \ \left( \begin{array}{cc}                
q^H + q^{-1} \la^2 F E \quad & q^{-1} \la F \\
\la q^{-H} E & q^{-H}
\end{array} \right), \qquad L \in \{ 0 , 1 \},
\end{equation}
where now
\beq\label{K2}
\cK = e^{- \pi i H}.                                                         
\eeq
The Casimir yields
\beq\label{m-rp}
\cK^L \act \gt c_1= \cK^L \act \frac{1}{ [2]_q} \tr_q ( \M ) = \cK^L 
\act M^4 \xrightarrow{\cB_{\La}} r   
e^{- \pi i L} \frac{\cos \frac{(\La + 1) \pi}{\k_1}}{\cos 
\frac{\pi}{\k_1}} \1. 
\end{equation}
Thus the quantum $\bZ_2$-\autn \eqref{out-m} reads                        
\beq
\eta : M_{\mu} \too - M_{\mu}, \quad \mu \in \{ 4, 0, 1, -1 \},
\end{equation}
just as required for the \aut to have the interpretation of an antipodal 
map on $\S_q^3$. The monodromy operator \eqref{K2} acts trivially only on 
integer-spinned \irrepsn.                                                   

We are now in a position to explicitly construct the q-matrix
model for the $\bZ_2^{antipod.}$-orbifold of the $\gt{su}_{\k}(2)$ WZW model
at $\k \in 4 \bN^{*}$.

\subsection{The orbifold.}

In the light of the general results of Sect.\ref{sub:fpr}. the coordinate
algebra of the $\bZ_2^{antipod.}$-orbifold of the stringy manifold
$\S^3_q$ is the algebra of quadratic monomials in the generators $M_{\mu}$
of $\resu2$, $\bZ_2$-extended - in the case of $\k \in 4 \bN^*$ - at the
unique central fixed point, $\La_{FP} = \tfrac{\k}{2}$. The algebra of
$\bZ_2$-invariants, defining a quantum manifold which could be called the
real quantum projective $3$-plane, $\bR P^3_q$, is easily verified to be
generated by the mutually independent operators:
\beq\label{orp3}
\cO \lb \bR P^3_q \rb = \spun \left\langle \1, M_0^2, M_1^2, M_{-1}^2,
M_0 M_1, M_{-1} M_0, M_4^2, M_4 M_0, M_4 M_1, M_4 M_{-1} \right\rangle / 
I_{RE,\det_q},
\end{equation}
where $I_{RE,\det_q}$ is the ideal defined by a set of relations, deriving 
directly from \eqref{req} and \eqref{recom}. The relations are not very 
illuminating and shall therefore be left out.

An important feature of the ensuing algebra is the central character of
$M_4^2$ which shall consequently be used to label inequivalent \irrepsn. Thus 
for $M_4^2 \neq 0$ (or $\La \neq \tfrac{\k}{2}$) we recover quantum $2$-spheres, 
as discussed in Sect.\ref{sub:q-orb-alg} and indicated by the BCFT. At the fixed
point, on the other hand, where the original RE and \eqref{req} simplify,
\beqa
M_0 M_{-1} = q^2 M_{-1} M_0 \quad &,& \quad
M_1 M_0 = q^2 M_0 M_1, \non \non
M_1 M_{-1} - M_{-1} M_1 = \la M_0^2 \quad &,& \quad
M_0^2 - q^{-1} M_1 M_{-1} - q M_{-1} M_1 \must r^2 \1,
\eeqa
the orbifold algebra reduces - upon choosing the rescaled generators:
\beq\label{oeqorb}
( P , R , \bar R , T , \bar T ) := \frac{[2]_q}{q^2 r^2} \lb \frac{1}{[2]_q}
M^2_0 , q^2 M^2_1 , q^2 M^2_{-1} , \frac{q}{\sqrt{[2]_q}} M_0 M_1 , -
\frac{q}{\sqrt{[2]_q}} M_{-1} M_0 \rb,
\end{equation}
to the $\bR P^2_q $ algebra found  in \cite{rp2}, generalized here to $q$ a 
phase and no longer real. According to the general discussion of 
Sec.\ref{sub:fpr} we have two inequivalent $\bR P^2_q $-branes corresponding to 
the two inequivalent \irreps of the stabiliser group $\bZ_2$ (with $D=1$ of 
Sec.\ref{sub:fpr} here).

We remark at this point that - as follows from the restriction to \irreps of 
vanishing monodromy (i.e. - in the case at hand - integer spin), effected by 
\eqref{K-mon} - it is only the integer spin branes that compose $RP^3_q$.

\paragraph{Space of functions}
~\nl
\indent Here we perform more scrupulous a comparison of the two descriptions 
of the fixed point geometry: the BCFT and the algebraic one. The modified 
(according to App.\ref{app:qBCFT}) fixed point BCFT brane geometry decomposes as 
($\k \in 4 \bN^*$):
\beq
\spun \left\langle \psi^{[\La_{FP}]_{\pm} [\La_{FP}]_{\pm}}_{I,i;0} 
\right\rangle_{I \in \ovl{0,\tfrac{\k}{2}} \cap 2\bN, \ i \in \ovl{0,2I}} \equiv 
\bigoplus_{\tfrac{K}{2} = 0}^{\tfrac{\k}{4}} (2K + 1).
\end{equation}
In the matrix model, on the other hand, we are free to take
arbitrary monomials in the generators $P,R,\bar R,T,\bar T$, with the
(even integer) spin of any such monomial determined by its overall
degree in the original coordinates $M_j, \ j \in \{ 0, 1, -1 \}$.
Restricting to the basis monomials (\cite{rp2}), ordered according to their
spin $2 s \in 2 \bN$,
\beq\label{bazar}
P^m R^{s-m}, \ P^n \bar{R}^{s-n}, \ P^l T R^{s-n-1}, \ P^n \bar T
\bar{R}^{s-n-1}, \qquad m \in \ovl{0,s}, \ n \in \ovl{0,s-1},
\end{equation}
we should obtain as many independent operators as there are even-spinned
primaries, which - clearly - is not the case\footnote{The origin of the
mismatch traces back to the failure of the original Drinfel'd twisting
procedure to render the ensuing $q$-deformed OPE algebra associative in
the entire range of spin parameters of the theory. It turns out that -
as was remarked already in \cite{ps} - the associativity breaks down as
early as $\La = E \lb \tfrac{\k}{4} \rb$, half-way between the north pole
of $\S^3_q$ and its equator.} unless we truncate the spin of the monomial
generators of the equatorial $\bR P^2_q$ algebra as $s \leq \frac{\k}{4}$.
The dimension of the space of monomials reduces accordingly,
\beq
\sum_{s = 0}^{S > \frac{\k}{4}} ( 4 s + 1 ) \too \sum_{s = 0}^{\frac{\k}{4}}  
( 4 s + 1 ) = \frac{1}{2} \lb \frac{\k}{2} + 1 \rb \lb \frac{\k}{2} + 2 \rb,
\end{equation}
to yield precisely the BCFT-dictated number of geometric degrees of freedom 
on the side of the matrix model. With this last observation we conclude our 
exposition of the simple example of quantum orbifold geometry.

\section{Summary.}\label{sum-out}

In the present paper, we advanced a quantum-algebraic model for curved 
non-com\-mu\-ta\-ti\-ve D-brane geometries on simple current orbifolds of the 
$SU(N+1)$ WZW manifolds. Following the original ideas of \cite{ps} we were able 
to consistently encode the BCFT data, regarded here as the ultimate foundation of 
all our constructions, in a simple algebraic framework based on the central idea 
of quantum group symmetry, as suggested by the structure of the CFT itself. Our 
study of the relevant Reflection Equation Algebras revealed significant structural 
similarities between the latter and the BCFT of the corresponding WZW 
models. The similarities, present both at the level of the respective \rep theories 
($\Out \lb A^{(1)}_N \rb$ versus $\Out \lb \reAN \rb$) and at the level of the 
algebra (the (resolved) orbifold BA versus the (cross-product-extended) 
$\reAN/\bZ_{N+1}$), provide nontrivial evidence for a close relationship between 
the $\qAN$-related algebras $\reAN$ and the stringy geometries defined by the BCFT. 
An important feature of this relationship is its naturalness. It follows - in 
particular - from the clear geometric meaning assigned to $\reAN$, which - in turn - 
induces a simple realisation of classical-type symmetries (whether continuous or 
discrete as in the present context) of the non-commutative manifolds defined by 
$\reAN$. One of the important manifestations of the relationship is a straightforward 
identification of the CFT monodromy charge in the quantum-algebraic setup. Last but 
not least, the explicit physical results on tensions of orbifold branes, falling in 
perfect agreement with the BCFT data, lend further support to our choice of the 
algebraic structure deforming the underlying BCFT\footnote{It is perhaps worth 
mentioning that - in addition to the calculations carried out explicitly in the paper 
- we are able within the present framework to test the quantum stability of the brane 
configurations \eqref{brane} as well as examine the ensuing inter-brane excitations 
using the techniques developed in \cite{qdyn}. So far our results seem to indicate 
towards stability against decay and reproduce an asymptotically correct picture of 
the lightest open strings stretched between distant branes.}.     

On the more formal side, we draw the reader's attention to the attractive pattern 
in the \rep theory of $\reAN$ uncovered in this paper, admitting a straightforward 
explanation in reference to the purely geometrical symmetries of the associated 
affine structure $A^{(1)}_N$. An immediate consequence of its presence is the 
construction of an entire class of new quantum geometries, wrapped by the 
fractional branes of the matrix model.  

Altogether, the arguments in favour of the proposal \eqref{brane} and 
\eqref{nactAN} and the associated orbifolding scheme are purely formal as well as 
physical in nature, turning our description of both the pre- and post-orbifolding 
$q$-geometry into a viable candidate for a quantum matrix model of (simple current 
orbifold) WZW geometry.\newline

\large{\textbf{Acknowledgements:}}

We would like to thank H.Steinacker and S.Watamura
for useful discussions. J.P. would also like to thank Satoshi Watamura for
his kind hospitality during J.P.'s visit at the Tohoku University.\newpage

\appendix
\noindent \large{\textbf{Appendices.}}

\section{The BCFT and its quantum deformation.}\label{app:qBCFT}

\subsection{The orbifold OPE algebra.}\label{sub:orb-OPE}

The present paper focuses on the study of a class of quantum algebras and
associated matrix models with the aim of encoding in them the physical content
of stringy WZW orbifolds. It thus seems natural to begin our discussion with
an exposition of some elements of the BCFT of simple current orbifolds
relevant to the subsequent quantum algebraic analysis.

The orbifolding procedure for WZW models was laid out in \cite{matsu}, which
we follow closely in this preparatory part, and leads to a formulation of
string theory on quotient spaces $G/\G$, with $\G$ - a subgroup of the
discrete group $\sOut$ of strictly affine\footnote{We give the name to
  the outer \auts of $\kmg$ which are not \auts of the horizontal algebra 
$\ggt$.} \auts of $\kmg$. It consists in dividing out the action of $\G$ which 
- at the level of the relevant OPE - is generated by simple currents, i.e. 
primaries with simple fusion rules with all other primaries. Denoting by 
$\La_g$ the weight label of a simple current corresponding to $g \in \G$, we 
have a fusion rule:
\beq
\exists !_{g\La \in \faff} \ : \ \La_g \x_{\mathcal{F}} \La = g\La
\end{equation}
for an arbitrary weight label $\La \in \faff$. The Abelian group formed by 
simple currents under fusion is known to be isomorphic with the group of 
strictly affine \autsn.
 
The action of $\G$ on the set of all primaries, labelled by weights
$\La \in \faff$, decomposes the latter into orbits and we take $[\La]$ to
label all boundary conditions associated with $\La$ by that action. Some of
the orbits may have fixed points, i.e. there may be weights stabilised by
subgroups of $\G$ for which we reserve the symbol $\cS_{\La} \subseteq \G$.
Among these there is a distinguished class of maximally stabilised weights
with $\cS_{\La} = \G$. Specialising to the case $\ggt = A_N$ and $\G = \sOut$
we shall call the corresponding fixed points central. An important property
of the stabiliser subgroups $\cS_{\La}$, used in the general BCFT
construction, is their independence of the choice of a particular representative 
of the orbit $[\La]$.
 
Another ingredient in the orbifolding recipe is the simple current
charge ${\Hat Q}_g (L) \in \bR/2\bZ$ of a given primary field $\Psi^{\La_1
\La_2}_{L,i}$ \wrt the simple current corresponding to $g$, determined by the
so-called braiding matrix\footnote{Cp \cite{sei-moore}.} of the underlying CFT,
\beq
(-1)^{{\Hat Q}_g (L)} := B^{(+)}_{L \La_g} \left[ \begin{smallmatrix} L &
\La_g \\ 0 & g\La \end{smallmatrix} \right].
\end{equation}
A closely related object is the monodromy charge:
\beq
Q_g (L) := {\Hat Q}_g (L) \mod 1 \quad \Longrightarrow \quad Q_g (L) = h_L +
h_{\La_g}- h_{gL} \mod 1,
\end{equation}
constant on simple current orbits for $h_{\La_g} \in \bZ$, in which case its
vanishing on $[\La]$ for all $g \in \G$ places the orbit among those to
survive the $\G$-orbifolding (\cite{yank-schenk}).

The last piece of the BCFT machinery we need to deal with orbifold
D-branes, in particular the fractional ones (\cite{fracD}), is a little group
theory of the
stabilisers. Indeed, according to the general theory we should have a unique
D-brane species over $\La$ for any of the inequivalent one-dimensional \irreps
of $\cS_{[\La]}$. We thus label the boundary states of the orbifold theory
with the corresponding characters $e_a : \cS_{[\La]} \to U(1)$. Upon
introducing the numbers:
\beq
d^{\; a \ L}_b := \frac{1}{\vert \cS_{[\La_1]} \cap \cS_{[\La_2]} \vert}
\sum_{h \in \cS_{[\La_1]} \cap \cS_{[\La_2]}} e_a (h) (-1)^{{\Hat Q}_h(L)}
e_b (h^{-1})
\end{equation}
for any pair of overlapping stabilisers $\cS_{[\La_1]} \cap \cS_{[\La_2]}
\neq \emptyset$ and $L$ such that there exists a non-zero fusion rule:
$N^{\; g\La_2}_{\La_1 \ L} \neq 0$ for some $g \in \G$, we then obtain the
partition functions for stabiliser-resolved orbifold D-branes ($\tau$ is the
standard modular parameter and $\cdot$ stands for the element-wise product
of groups):
\beq\label{partf-res}
Z^{orb}_{[\La_1]_a [\La_2]_b} (\tau) = \frac{1}{\vert \cS_{[\La_1]} \cdot
\cS_{[\La_2]} \vert} \sum_{g \in \G} \sum_{L \in \faff} N_{\La_1 \ L}^{\;
g\La_2} d^{\; a \ L}_b \chi_L (\tau),
\end{equation}
summing up to the partition function of orbifold orbits (or unresolved
D-branes):
\beq
Z^{orb}_{[\La_1] [\La_2]} (\tau) = \sum_{g \in \G} \sum_{L \in \faff}
N_{\La_1 \ L}^{\; g\La_2} \chi_L (\tau).
\end{equation}
From \eqref{partf-res} we now readily derive the tensions of the fractional
branes\footnote{Cp \cite{cou}.} by specialising the formula to the case 
$\cS_{\La_2} = \{ \id \}$ when it becomes
\beq\label{frac-ten-bcft}
Z^{orb}_{[\La_1]_a [\La_2]} (\tau) = \frac{1}{\vert \cS_{[\La_1]} \vert}
\sum_{g \in \G} \sum_{L \in \faff} N_{\La_1 \ L}^{\; g\La_2} \chi_L (\tau) =
\frac{1}{\vert \cS_{[\La_1]} \vert} Z^{orb}_{[\La_1] [\La_2]} (\tau).
\end{equation}
Hence the graviton coupling between the fractional D-brane carrying an
arbitrary stabiliser label $a$ associated with $\cS_{\La_1}$ and an
off-fixed-point one contributes the fraction of $\tfrac{1}{\vert
\cS_{[\La_1]} \vert}$ to the overall graviton coupling between the
(unresolved) D-branes, an intuitive result we shall demonstrate to be
reproduced by the matrix model of Sect.\ref{q-orb-mm}.

Having introduced all the relevant formal instruments we may now
define an action of $\G$ on the primaries of the pre-orbifolding theory:
\beq\label{def-act}
g \act \Psi^{\La_1 \La_2}_{L,i} (x) := (-1)^{- {\Hat Q}_g(L)} \Psi^{g\La_1
g\La_2}_{L,i} (x),
\end{equation}
easily verified to be consistent with the boundary OPE \eqref{res-OPE} and 
its off-fixed-point version due to the following property of the fusing 
matrix (\cite{matsu}):
\beq \label{scale}
\forall_{g \in \G} \ : \ F_{gJ,k} \left[ \begin{smallmatrix} i & j \\ gI &
gK \end{smallmatrix}
\right]^{\a_i,\a_j;\a_k}_{\k} = (-1)^{{\Hat Q}_g (i) + {\Hat Q}_g (j) -
{\Hat Q}_g (k)} F_{J,k} \left[ \begin{smallmatrix} i & j \\ I
& K \end{smallmatrix} \right]^{\a_i,\a_j;\a_k}_{\k}.
\end{equation}
The definition \eqref{def-act} provides us with a possibility to average
over $\G$ primaries interpolating between off-fixed-point boundary states,
whereby the associated orbifold primaries are obtained,
\beq\label{off-prim}
\Psi^{[\La_1] [\La_2]}_{L,i;g} (x) := \sum_{g' \in \G} g' \act \Psi^{\La_1
g\La_2}_{L,i} (x).
\end{equation}
Supplementing the above formula with its fixed-point counterpart\footnote{A
definition proved sensible in \cite{matsu}.}:
\beq\label{fp-prim}
\Psi^{[\La_1]_a [\La_2]_b}_{L,i;g} := \sum_{g_1 \in \cS_{\La_1}} \sum_{g_2
\in \cS_{\La_2}} \Psi^{[\La_1] [\La_2]}_{L,i;g_1 g g_2^{-1}} (-1)^{-
{\Hat Q}_{g_1}(L)} e_a(g_1) e_b(g_2^{-1}),
\end{equation}
with the simple current label $g$ in both \eqref{off-prim} and \eqref{fp-prim}
such that there is a non-zero fusion rule: $N^{\; g\La_2}_{\La_1 \ L} \neq 0$,
we may finally write down the OPE of the stabiliser-resolved boundary
primaries:
\bal
& \Psi^{[\La_1]_{a_1} [\La_2]_{a_2}}_{L_1,i_1;g_{12}} (x_1)
\Psi^{[\La_2]_{a_3} [\La_3]_{a_4}}_{L_2,i_2;g_{23}} (x_2) =
\delta_{a_2,a_3} \sum_{g \in \cS_{\La_2}} \sum_{L_3 \in \faff}
\sum_{i_3 = 1}^{N^{\La_3}_{\La_1 L_3}} x_{12}^{h_1 + h_2 - h_3}
\Psi^{[\La_1]_{a_1} [\La_3]_{a_4}}_{L_3,i_3;g_{123}g} (x_2) \x \non \non
& \x (-1)^{- {\Hat Q}_{g_{12}g}(L_2)} e_b(g) F_{g_{12}\La_2,L_3} \bigl[
\begin{smallmatrix} L_1 & L_2 \\ \La_1 & g_{123}g\La_3
\end{smallmatrix} \bigr]^{i_1,i_2;i_3}_{\k} c^{L_1 L_2 L_3}_{i_1 i_2 i_3},
\label{res-OPE}
\end{align}
where we have used the shorthand notation: $x_{12} := x_1-x_2$ and $g_{123} 
:= g_{12} g_{23}$. An analogous formula for off-fixed-point boundary states 
can be obtained from \eqref{res-OPE} by taking trivial stabiliser labels.

Prior to passing to the $q$-deformed OPE algebra we make, after
\cite{matsu}, one more significant remark: according to \cite{geowzw} the
OPE algebra of boundary primaries is a stringy deformation of the associative
algebra of functions on the target geometry; the emergence of stabiliser
resolution and the introduction of charge-weighted averages over $\G$ in
the above OPE has - in this spirit - been considered to reflect the
existence of an algebraic structure called the crossed product extension
of the algebra of functions on the orbifold, present in the known matrix
models of fixed-point geometries (\cite{orb-gen}). We shall have more to say
about this issue in Sect.\ref{q-orb-mm}.

\subsection{The monodromy projection.}\label{sub:mono}
                        
There remains one more essential element of the CFT orbifolding procedure that 
we have not considered in detail so far, namely: 
restriction (in the orbifold theory) to the boundary states with a trivial 
monodromy charge. It is imposed, in particular, on the relevant partition 
function for closed strings on the group manifold $G$ as
\beq\label{transorb}
Z_G (\tau) \too Z_{G/\G} (\tau) = Z^{\mbox{\tiny untwisted}}_{G/\G} 
(\tau) + \sum_{k = 1}^{\vert \G \vert} \g^k \act 
Z^{\textrm{\tiny untwisted}}_{G/\G} (\tau), 
\end{equation}    
with the action of the single generator $\g$ of the subgroup $\G$ of the simple 
current group\footnote{We are going to work with $SU(N+1)$ groups in the sequel, 
hence the particular choice of the rank of the simple current group.} $\G_{N+1} 
\cong \bZ_{N+1}$ on the standard character bilinears:
\beq\label{untwist}
Z^{\textrm{\tiny untwisted}}_{G/\G} (\tau) = 
\sum_{\La \in \faff,\  Q_{\g} (\La) = 0} 
\chi_{\La} (\tau) \chi^{*}_{\La} 
(\tau) 
\end{equation}  
defined as
\beq
\g \act \chi_{\La} (\tau) \chi^{*}_{\La} (\tau) := \chi_{\g \La} (\tau) 
\chi^{*}_{\La} (\tau),
\end{equation}
where $\g \La$ is the unique dominant integral affine weight assigned to $\La \in 
\faff$ by the simple current $\g$ and $Q_{\g}(\La)$ is - as earlier - the
monodromy charge of the weight $\La$ \wrt the generating current $\g$. 

The transition \eqref{transorb} is, in fact, an instance of averaging - in the 
spirit of \cite{dix} - \wrt the action of the orbifold group at the level of $G$. 
We may readily convince ourselves of the validity of that statement by making use 
of the crucial formula (\cite{orient}):
\beq\label{mono-exp}
Q_{\g_{N+1}}(\La) = \frac{\cC (\La)}{N+1} \mod 1
\end{equation}
establishing a simple relation between the monodromy charge of $\La$ \wrt the 
generator $\g_{N+1}$ of the full simple current group, $\bZ_{N+1}$, and the so-called 
congruence $\cC (\La)$ of the weight $\La$. The latter is a function on $P^* (\ggt) / 
P (\ggt)$ (with $P (\ggt)$ - the root space of $\ggt$),
a space isomorphic to the centre $Z (G)$ of the group $G$, which - in turn - 
coincides with $\bZ_{N+1}$ (cp \cite{fuchs-book}). Upon choosing the Chevalley basis 
for $\ggt$ we may take the generator $c_{N+1}$ of $Z (G)$ such ($\nu$ is the so-called 
congruence vector of $\ggt$)
\beq\label{monod}
c_{N+1} = e^{\frac{2 \pi i}{N+1} \sum_{i = 1}^N \nu_i H_i}, \quad \nu := 
( 1 ,   2 , \ldots , N )  
\end{equation} 
that in restriction to the \irrep of the weight $\La$ it yields\footnote{It
  ought to be emphasised that $\cC$ is well-defined on \irrepsn, it being a
  function  on $P^* (\ggt) / P (\ggt)$.}: 
\beq
c_{N+1} \vert_{R_{\La}} = e^{\frac{2 \pi i \cC (\La)}{N+1}} = e^{2 \pi i 
Q_{\g_{N+1}}(\La)}.
\end{equation}  
The last identity enables us to rewrite \eqref{untwist} in the following manner:
\beqa
Z^{\textrm{\tiny untwisted}}_{G/\G} (\tau) &=& \frac{1}{\vert \G \vert} 
\sum_{\g \in \G} \sum_{\La \in \faff} R_{\La} ( \g ) \chi_{\La} (\tau) 
\chi^{*}_{\La} (\tau) = \non \non
&=& \frac{1}{\vert \G \vert} \sum_{k = 0}^{\vert \G \vert - 1} 
\sum_{\La \in \faff} R_{\La} ( c^{d \cdot k}_{N+1} ) \chi_{\La} (\tau) 
\chi^{*}_{\La} (\tau), \quad  d := \frac{N+1}{\vert \G \vert} \overset{!}{\in} \bN, 
\label{untwist-centr}
\eeqa
which completes our demonstration. The above averaging was employed quite 
explicitly in \cite{gep-witt} to obtain modular invariants for the orbifolds 
$SU(2)/\bZ_2$ and $SU(3)/\bZ_3$.

In conclusion, we note that \eqref{untwist-centr} provides a clear geometric 
interpretation of the projection \eqref{untwist} onto the subset of \irreps of 
$\ggt$ carrying the trivial monodromy charge, namely: it should be understood as a 
transition from the $(N+1)$-element set\footnote{The number $N+1$ is known in 
the present context as the index of connection of $G$.} of congruence classes 
of adjoint orbits (or conjugacy classes, labelled by the corresponding weights) 
within the group manifold of $G$ to the $d$-element subset consisting of all 
congruence classes associated with the unique non-faithful (trivial) \irrep of 
$\G \subset Z (G)$. The transition, on the other hand, follows from dividing 
out the action of simple currents which - at the level of $G$ - become encoded 
in $Z (G)$. 
     
It seems well worth remarking that the averaging over the subset $\G$ of the 
centre $Z (G)$ of the group descends from the somewhat abstract level of 
partition functions to the level of the orbifold boundary OPE algebras. Indeed, 
a closer look at \eqref{off-prim} and \eqref{def-act} reveals the presence of 
the familiar structure\footnote{We restrict ourselves to the off-fixed-point 
case for clarity of the exposition only.}:
\beq
\Psi^{[\La] [\La]}_{L,i;\g} (x) = \sum_{\g' \in \G \subset Z (G)} R_L ( \g' ) 
\Psi^{\g' \La \; \g' \g \La}_{L,i} (x).
\end{equation}
Given the geometric interpretation of both the projection and the boundary OPE 
algebras themselves it is quite natural to expect the above general pattern to 
repeat itself in the quantum-algebraic setup to be developed.

\section{Quantum deformation.}\label{sec:deform}

The first step towards an algebraic description of WZW D-branes is taken
at the level of the general boundary OPE:
\beq\label{b-OPE}
\Psi^{\La_1 \La_2}_{L_1,i_1} (x_1) \Psi^{\La_2 \La_3}_{L_2,i_2} (x_2)
\sim_{\tx{\ciut{OPE}}} \sum_{L_3 \in \faff} \sum_{i_3 = 1}^{N^{\La_3}_{\La_1
\ L_3}} x_{12}^{h_1 + h_2 - h_3} \Psi^{\La_1 \La_3}_{L_3,i_3} (x_2)
F_{\La_2 L_3} \bigl[ \begin{smallmatrix} L_1 & L_2 \\ \La_1 & \La_3
\end{smallmatrix} \bigr]^{i_1,i_2;i_3}_{\k} c^{L_1 L_2 L_3}_{i_1 i_2 i_3}, 
\eeq
in which $F_{\La_2 L_3} \bigl[ \begin{smallmatrix} L_1 & L_2 \\ \La_1 & \La_3 
\end{smallmatrix} \bigr]^{i_1,i_2;i_3}_{\k}$ is the fusion matrix of the model 
and $c^{L_1 L_2 L_3}_{i_1 i_2 i_3}$ are structure constants encoding the 
group-theoretic nature of the indices carried by the horizontal descendants 
$\Psi^{\La_1 \La_2}_{L_1,i_1}$ of boundary primary fields $\Psi^{\La_1 
\La_2}_{L_1}$ (for details cp, e.g., \cite{schom}).

With the aim of extracting from \eqref{b-OPE} the D-brane geometry we make the 
usual assignment (cp \cite{mat-wzw} and earlier papers on non-commutative 
stringy geometries):
\beq\label{trans}
\Psi^{\La_1 \La_2}_{L_1,i_1} (x_1) \too \psi^{\La_1 \La_2}_{L_1,i_1},
\end{equation}
so that the algebraic content of \eqref{b-OPE} is preserved,
\beq\label{b-alg}
\psi^{\La_1 \La_2}_{L_1,i_1} \star \psi^{\La_2 \La_3}_{L_2,i_2} := \sum_{L_3
\in \faff} \sum_{i_3 = 1}^{N^{\La_3}_{\La_1 \ L_3}}
\psi^{\La_1 \La_3}_{L_3,i_3} F_{\La_2 L_3} \bigl[ \begin{smallmatrix} L_1 &
L_2 \\ \La_1 & \La_3 \end{smallmatrix} \bigr]^{i_1,i_2;i_3}_{\k} c^{L_1 L_2
L_3}_{i_1 i_2 i_3},
\end{equation}
and only the world-sheet dependence is dropped\footnote{One can motivate this       
transition by considering the $\k \to \infty$ limit of \eqref{b-OPE} for a set 
of boundary labels $\La_i$ sufficiently close to the trivial weight.}. The 
Boundary Algebra (BA) thus defined is readily demonstrated to be non-associative 
and so it has to be deformed to be embedded in a matrix algebra. The 
non-associativity follows from its hybrid quantum-classical structure. Indeed, 
while the three-point structure constants $c^{L_1 L_2 L_3}_{i_1 i_2 i_3}$ are 
classical intertwiners (in the simplest case of the $\gt{su}_{\k} (2)$ model 
they are just the ordinary Clebsch--Gordan coefficients of $SU(2)$) the fusion 
matrix is already a quantum entity\footnote{The matrix enters the BCFT analysis 
at a rather abstract stage and both its r\^ole and uniqueness - a consequence 
of strict algebraic constraints imposed upon it - determine it as an already
quantum(-group-theoretic) object. Cp \cite{sei-moore,qg-cft,mooresh}.}.
The last observation leads us to the idea of deforming \eqref{b-alg},
\beq\label{ass}
c^{L_1 L_2 L_3}_{i_1 i_2 i_3} \too \tilde{c}^{L_1 L_2 L_3}_{i_1 i_2 i_3},
\end{equation}
in such a way:
\bgt
\sum_{l,\a_l} F_{J,l} \left[ \begin{smallmatrix} i & j \\ I & K
\end{smallmatrix}
\right]^{\a_i,\a_j;\a_l}_{\k} F_{K,m} \left[ \begin{smallmatrix} l & k \\
I & L \end{smallmatrix} \right]^{\a_l,\a_k;\a_m}_{\k} \tilde{c}^{\a_i \a_j
\a_l}_{i j l} \tilde{c}^{\a_l \a_k \a_m}_{l k m} = \non \non
= \sum_{l,\a_l} F_{J,m} \left[ \begin{smallmatrix} i & l \\ I & L
\end{smallmatrix} \right]^{\a_i,\a_l;\a_m}_{\k} F_{K,l} \left[
\begin{smallmatrix} j & k \\ J & L \end{smallmatrix}
\right]^{\a_j,\a_k;\a_l}_{\k} \tilde{c}^{\a_j \a_k \a_l}_{j k l}
\tilde{c}^{\a_i \a_l \a_m}_{i l m} \label{Fc-rel}
\end{gather}
as to turn the latter purely quantum and associative\footnote{At least
within a certain (truncated) range of its group-theoretic labels,
cp \cite{ps}.}, so that
\beq\label{pre-assoc}
\lb \psi^{\La_1 \La_2}_{L_1,i_1} \star_q \psi^{\La_2 \La_3}_{L_2,i_2} \rb
\star_q \psi^{\La_3 \La_4}_{L_3,i_3} = \psi^{\La_1 \La_2}_{L_1,i_1}
\star_q \lb \psi^{\La_2 \La_3}_{L_2,i_2} \star_q \psi^{\La_3 \La_4}_{L_3,i_3}
\rb.
\end{equation}
obtains for $\star_q$ defined as $\star$ in \eqref{b-alg} but with the
substitution \eqref{ass}. In the above-mentioned $\gt{su}_{\k} (2)$ case
the deformation boils down to replacing the classical Clebsch--Gordan
coefficients with those of $\qU2$ and was given in \cite{mat-wzw}, where it
first appeared, an interpretation in terms of the so-called Drinfel'd twist.
The idea was developed and exploited in \cite{ps}, with essential emphasis
on the indication, contained in the new boundary algebra, towards a quantum
symmetry of the underlying D-brane geometry and, consequently, of an
associated matrix model, built on the assumption of quantum group covariance.
Indeed, by adducing the standard Wigner--Eckhart argument, we identify the    
new boundary operators $\psi^{\La_1 \La_2}_{L_1,i_1}$ with certain 
$\cU_q(\gt{g})$-intertwiners\footnote{Cp \cite{fuzzy}.}, furnishing an adjoint 
module of the quantum algebra. A transition from the left-right 
$\cU_q(\gt{g})$-symmetry of the ``bulk'' to its thus encoded vector part lies 
at the heart of the quantum-algebraic framework presented.                    
It is in this sense that the new algebra constitutes the basis, on the BCFT 
side, of the models developed in \cite{ps}.

There is yet another feature of \eqref{ass} which becomes particularly
significant in our present context. The deformation \eqref{ass}-\eqref{Fc-rel}
proves sufficient to turn the orbifold analogue of \eqref{b-alg} associative.
The proof of the last statement is presented below and enables us to start the 
construction of the matrix model of the orbifold physics directly at the level 
of the original quantum matrix algebra of \cite{ps} and seek for the \auts of 
the latter corresponding to the elements of the simple current orbifold group. 
The ensuing quotient structure is expected to define the quantum geometry of 
untwisted D-branes wrapping the orbifold.

\subsection{Proof of the associativity of the deformed boundary
algebra.}\label{app:assoc}

In this appendix we explicitly prove the useful fact: given \eqref{pre-assoc},
the deformation of the orbifold Boundary Algebra,
\bal
& \psi^{[\La_1]_{a_1} [\La_2]_{a_2}}_{L_1,i_1;g_1} \star_q \psi^{[\La_2]_{a_3}
[\La_3]_{a_4}}_{L_2,i_2;g_2} = \non \non
& = \d_{a_2,a_3} \sum_{g \in \cS_{\La_2}} \sum_{L_3,i_3} \psi^{[\La_1]_{a_1}
[\La_3]_{a_4}}_{L_3,i_3;gg_{12}} e_{a_2} (g) (-1)^{- {\Hat Q}_{gg_1}(L_2)}
F_{g_1\La_2,L_3} \left[ \begin{smallmatrix} L_1 & L_2 \\ \La_1 & gg_{12}\La_3
\end{smallmatrix} \right]^{i_1,i_2;i_3}_{\k} \tilde{c}^{L_1 L_2 L_3}_{i_1 i_2
i_3}, \label{orb-q-balg}
\end{align}
renders the latter associative (we retrieve the unresolved case from
\eqref{orb-q-balg} upon setting $g = \id = g'$):
\bal
& \lb \psi^{[\La_1]_{a_1} [\La_2]_{a_2}}_{L_1,i_1;g_1} \star_q
\psi^{[\La_2]_{a_3} [\La_3]_{a_4}}_{L_2,i_2;g_2} \rb \star_q
\psi^{[\La_3]_{a_5} [\La_4]_{a_6}}_{L_3,i_3;g_3} = \non \non
& = \delta_{a_2,a_3} \sum_{g \in \cS_{\La_2}} \sum_{L_4, i_4}
\psi^{[\La_1]_{a_1} [\La_3]_{a_4}}_{L_4,i_4;gg_{12}}
\star_q \psi^{[\La_3]_{a_5} [\La_4]_{a_6}}_{L_3,i_3;g_3} e_{a_2} (g)
(-1)^{- {\Hat Q}_{gg_1}(L_2)} \x \non \non 
& \x F_{g_1\La_2,L_4} \left[ \begin{smallmatrix} L_1 & L_2 \\ \La_1 & 
gg_{12}\La_3 \end{smallmatrix} \right]^{i_1,i_2;i_4}_{\k}
\tilde{c}^{L_1 L_2 L_4}_{i_1 i_2 i_4} = \non \non
& = \delta_{a_2,a_3} \delta_{a_4,a_5} \sum_{\substack{g \in \cS_{\La_2} \\
g' \in \cS_{\La_4}}}
\sum_{\substack{L_4,i_4 \\ L_5,i_5}} \psi^{[\La_1]_{a_1}
[\La_4]_{a_6}}_{L_5,i_5;g'gg_{123}} e_{a_4} (g') e_{a_2} (g)
(-1)^{- {\Hat Q}_{g'gg_{12}}(L_3) - {\Hat Q}_{gg_1}(L_2)} \x \non \non
& \x F_{gg_{12}\La_3,L_5} \left[ \begin{smallmatrix} L_4 & L_3 \\ \La_1 & 
g'gg_{123}\La_4 \end{smallmatrix} \right]^{i_4,i_3;i_5}_{\k} 
F_{g_1\La_2,L_4} \left[ \begin{smallmatrix} L_1 & L_2 \\ \La_1 &
gg_{12}\La_3 \end{smallmatrix} \right]^{i_1,i_2;i_4}_{\k}
\tilde{c}^{L_1 L_2 L_4}_{i_1 i_2 i_4} \tilde{c}^{L_4 L_3 L_5}_{i_4 i_3 i_5} =
\non \non
& = \delta_{a_2,a_3} \delta_{a_4,a_5} \sum_{\substack{g \in \cS_{\La_2} \\ g'
\in \cS_{\La_3}}} \sum_{\substack{L_4,i_4 \\ L_5,i_5}}
\psi^{[\La_1]_{a_1} [\La_4]_{a_6}}_{L_5,i_5;g'gg_{123}} e_{a_4} (g')
e_{a_2} (g) (-1)^{- {\Hat Q}_{gg_1}(L_4) - {\Hat Q}_{g'g_2}(L_3)} \x \non \non
& \x F_{g_2\La_3,L_4} \left[ \begin{smallmatrix} L_2 & L_3 \\ 
\La_2 & g'g_{23}\La_4 \end{smallmatrix} \right]^{i_2,i_3;i_4}_{\k} 
F_{g_1\La_2,L_5} \left[ \begin{smallmatrix} L_1 & L_4 \\ \La_1 & 
g'gg_{123}\La_4 \end{smallmatrix} \right]^{i_1,i_4;i_5}_{\k}
\tilde{c}^{L_1 L_2 L_4}_{i_1 i_2 i_4} \tilde{c}^{L_4 L_3 L_5}_{i_4 i_3 i_5} =
\nonumber
\end{align}
\bal
& = \delta_{a_4,a_5} \sum_{g' \in \cS_{\La_3}} \sum_{L_4,i_4}
\psi^{[\La_1]_{a_1} [\La_2]_{a_2}}_{L_1,i_1;g_1} \star_q \psi^{[\La_2]_{a_3}
[\La_4]_{a_6}}_{L_4,i_4,g'g_{23}} e_{a_4} (g') (-1)^{- {\Hat Q}_{g'g_2}(L_3)}
\x \non \non
& \x F_{g_2\La_3,L_4} \left[ \begin{smallmatrix} L_2 & L_3 \\ \La_2 & 
g'g_{23}\La_4 \end{smallmatrix} \right]^{i_2,i_3;i_4}_{\k}
\tilde{c}^{L_2 L_3 L_4}_{i_2 i_3 i_4} = \non \non
& = \psi^{[\La_1]_{a_1} [\La_2]_{a_2}}_{L_1,i_1;g_1} \star_q \lb
\psi^{[\La_2]_{a_3} [\La_3]_{a_4}}_{L_2,i_2;g_2} \star_q \psi^{[\La_3]_{a_5}
[\La_4]_{a_6}}_{L_3,i_3;g_3} \rb \square.
\end{align}

\section{The algebra $\mathbf\qANex$.}\label{sec:u-ext}

We begin with the definition of $\qANex$ which we take to be generated by the 
elements:
\beq\label{qanex}
k_{\pm \ep_i}, \quad i \in \overline{1,N+1} \quad ; \quad E_j , F_j, \quad j 
\in \overline{1,N},
\end{equation}
subject to the
relations \footnote{We denote $\la := q - q^{-1}$, $[2] := q +
q^{-1}$ and $k_{\pm n \ep_i} := k_{\pm \ep_i}^n,\;n \in \bN$}:
\bgt
\kp{i} \kp{j} = \kp{j} \kp{i} \quad , \quad \kp{i} \km{i} = \1 = \km{i}
\kp{i}, \label{DJ1} \\ \non
\kp{1} \kp{2} \cdots \kp{N+1} = \1, \label{DJ2} \\ \non
\kp{i} E_j \km{i} = q^{\d_{i j} - \d_{i-1,j}} E_j \quad , \quad
\kp{i} F_j \km{i} = q^{- \d_{i j} + \d_{i-1,j}} F_j, \label{DJ3} 
\end{gather}
\bgt
[ E_i , F_j ] = \d_{ij} \frac{\kp{i}\km{i+1} - \km{i}\kp{i+1}}{\la},
\label{DJ4} \\ \non
E_i^2 E_j - [2] E_i E_j E_i + E_j E_i^2 = 0 \quad , \quad F_i^2 F_j -
[2] F_i F_j F_i + F_j F_i^2 = 0 \quad \rm{for} \ \vert i - j \vert =
1, \label{DJ5} \\ \non
E_i E_j = E_j E_i \quad , \quad F_i F_j = F_j F_i \quad \rm{for} \
\vert i - j \vert > 1. \label{DJ6}
\end{gather}
The Cartan generators $k_{\pm \ep_j}, j \in \overline{1,N+1}$ are
defined in direct reference to the standard embedding of the root
space of the classical algebra, $P(A_N)$, in $\mathbb{R}^{N+1}$.
We thus have the well-known transformation between the simple-root basis,
$\a_i$, and the orthonormal (Cartesian) one, $\ep_i$,
\beq\label{epsi}
\a_i = \ep_i - \ep_{i+1}, \quad i \in \overline{1,N}, \qquad \lb \ep_i ,
\ep_j \rb = \d_{i j},
\end{equation}
with the orthogonality condition on weights $\La$: 
$\sum_{i = 1}^{N+1} \lb \La , \ep_i \rb \must 0$.
Clearly, the orthogonality condition is encoded in \eqref{DJ2} at the
level of the algebra.

The algebra is endowed with a Hopf structure. In particular, it has
the antipode:
\beq
S k_{\pm \ep_i} = k_{\pm \ep_i} \quad , \quad S E_i = - \km{i} \kp{i+1} E_i
\quad , \quad S F_i = - F_i \kp{i} \km{i+1},
\end{equation}
employed frequently in the sequel.

\subsection{The centre of $\qANex$.}\label{app:ex-centr}

Another interesting aspect of the general theory of the
extended quantum enveloping algebras is the structure of their
centre, $Z_q (A_N)$, playing a crucial r\^ole in any
representation-theoretic analysis. In the case of the deformation
parameter $q$ being the $2\k_N$-th primitive root of unity the centre is
known (\cite{arnba}) to be generated by the scalar operators:
\beq
Z_0 = \spun \left\lan k_{ \ep_i}^{2\k_N} , e^{2\k_N}_{i j} ,
f^{2\k_N}_{i j} \right\ran_{i,j \in \overline{1,N+1}}, \label{expc}
\end{equation}
peculiar to the root-of-unity case, and the Casimir (scalar) operators:
\beq
Z_1 = \spun \left\lan \cC_k \right\ran_{k \in \overline{1,N}},
\label{polyCas}
\end{equation}
explicitly given by \eqref{re-ce} through \eqref{re-to-u}. On the irreducible 
\hw \reps $e^{2\k_N}_{i j} = 0 = f^{2\k_N}_{i j}$.

\subsection{Relations between $\reAN, \ \cU_q^{ext}(A_N)$ and $\cU_h(A_N)$.}
\label{sec:rea}

There is a set of algebra homomorphisms:
\bgt
\cU_q(A_N) \to \cU_q^{ext}(A_N) \to \cU_h(A_N) \non
\uparrow\;\;\;\; \label{alg-hom} \\
\reAN \nonumber
\end{gather}
which we describe in the course of the paper. We denote the corresponding
generators as:
\bit
\item the quantum enveloping algebra $\qAN$:\\
$\{ K_j , K_j^{-1} , E_j , F_j \}_{j \in \overline{1,N}}$;
\item the extended quantum enveloping algebra $\qANex$:\\
$\{ k_{\pm \ep_i} , E_j , F_j \}_{i \in \overline{1,N+1},\ j \in
\overline{1,N}}; $
\item the $h$-adic Hopf algebra $\hAN$:\\
$\{ H_j , E_j , F_j \}_{j \in \overline{1,N}}$.
\end{itemize}
The only nontrivial images of the generators, given by the homomorphisms
\eqref{alg-hom}, are
\bgt
\forall_{j \in \overline{1,N}} \ : \ K_j \to k_{\ep_j}k_{-\ep_{j+1}},
\label{alg-map-q} \\ \non
\forall_{i \in \overline{1,N+1}} \ : \ k_{\ep_i} \to q^{H_{\ep_i}},
\label{alg-map-h}
\end{gather}
where ($\La_j$ are the fundamental weights)
\beq\label{hepj}
 H_{\ep_i} : = \sum_{j = 1}^N
(\ep_i , \La_j )H_j.
\end{equation}

\subsection{$\Rep(\qANex)$ and $\Out(\qANex)$.}\label{sub:rep-out-rea}

Automorphisms of $\qANex$ are of the phase-changing type\footnote{We do not    
consider the standard Dynkin diagram reflection.}:
\bgt
\lb k_{\pm \ep_i} , E_j , F_j \rb \too \lb e^{\pm i \pi p_i}
k_{\pm \ep_i} , E_j , e^{i \pi ( p_j - p_{j+1} )} F_j \rb, \quad
i \in \overline{1,N+1}, \ j \in \overline{1,N}, \label{phex} \\ \non
2( p_j - p_{j+1} ) = 0 \mod 2 \quad , \quad \sum_{l = 1}^{N+1}
p_l= 0 \mod 2. \label{conphex}
\end{gather}
As the homomorphisms \eqref{alg-hom} identify the ladder generators
($E_j,F_j$) of the algebras considered, their finite dimensional 
irreducible \hw modules are isomorphic and have the same \hws (denoted as 
$V_\La$). The only effect of the \auts \eqref{phex}-\eqref{conphex} is a 
change of the phases of the eigenvalues of the Cartan generators $k_{\pm 
\ep_i}$.

We need to determine the range of $p_i's$. According to the \rep theory of 
$\qAN$ (Chapter 7 of \cite{ks}) and \eqref{alg-map-q}-\eqref{alg-map-h},
\beqa\label{eigenkep}
 k_{\ep_j} k_{j+1}^{-1} \must e^{i \pi \om_j} q^{H_j}, \quad \om := \lb 
\om_1 , \om_2 , \ldots , \om_N \rb \in \bZ_2^N.
\eeqa
Imposing \eqref{DJ2} we further constrain the parameters in \eqref{eigenkep}:
\beqa
\forall_{j \in \overline{2,N+1}} \ : \ k_{\ep_j} &\must& e^{- i \pi 
\sum_{m = 1}^{j-1} \om_m} q^{- \sum_{n = 1}^{j-1} H_n} k_{\ep_1}, \non \\ 
k_{\ep_1}^{N+1} &\must& e^{i \pi \sum_{m = 1}^N ( N + 1 - m ) \om_m}
q^{\sum_{n = 1}^N ( N + 1 - n ) H_n}, \nonumber
\eeqa
which means that we have an $(N+1)$-tuple of \autsn:
\bgt
k_{\ep_1} \to e^{- \frac{i \pi L(l,\om)}{N+1}} k_{\ep_1} ,\non
 \\
\forall_{j \in \overline{2,N+1}} \ : \ k_{\ep_j}\to e^{- \frac{i \pi
L(l,\om)}{N+1}} e^{- i \pi \sum_{m = 1}^{j-1} \om_m} k_{\ep_j}, \label{espc-a}
\nonumber
\end{gather}
where
\beq
L(l,\om) = 2l+\sum_{m =
1}^N ( N + 1 - m ) \om_m,\quad l \in \bZ_{N+1}.
\end{equation}
The latter provide also the most general solution to \eqref{conphex} upon
a straightforward identification of phases $p_i$. The associated group $\Out 
(\qANex)$ is 
\beq \Out (\qANex) = \lb \bZ_2^N \ox
\bZ_{N+1} \rb \lx \bZ_2,
\label{outex}
\end{equation}
where the distinguished $\bZ_2$ factor corresponds to the classical mirror 
symmetry of the Dynkin diagram, present already in $\Out(\hAN)$ (\cite{ks}).

The irreducible \hw \reps $\cR^{l,\om}_\La$ of $\qANex$ are then of the form: 
\beq 
\cR^{l,\om}_\La \subset \Rep_{\tx{h.w.}} ( \qANex ) {\cong} \Rep_{\tx{r.h.w.}} 
( \hAN ) \ox \lb \bZ_{N+1} \ox \bZ_2^N \rb,
\end{equation}
as follows from \eqref{espc-a}. Beware: we do not claim that all of them are
inequivalent although it is certainly true for an arbitrary weight $\La \in 
\faffN$ such that all Casimir operators are non-zero on, say, $\cR^{0,0}_\La$.

\section{The algebra $\reAN$.}

In this paper we are interested in representations of  $\reAN$ induced by the
homomorphism:
\beqa\label{re-uqa}
\reAN \to \cU_q^{ext}(A_N). 
\eeqa
The homomorphism was originally discussed in
\cite{mooresh,resh-sts} (following the earlier results of \cite{frt}) and
reads
\beq\label{re-to-u}
{M} = \mathbf{L}^{+} S\mathbf{L}^{-}\in \Mat((N+1) \x (N+1) ;
\mathbb{C}) \ox \qANex.
\end{equation}
where
\beq
\mathbf{L}^{\pm} = \sum_{i,j = 1}^{N+1} \me_{ij} \ox L^{\pm}_{i j}
\end{equation}
are operator-valued matrices\footnote{$\me_{i j}$ are the basis matrices
$(\me_{i j})_{k l} = \d_{i k} \d_{j l}$.} presented explicitly below.  
Notice that with
\eqref{re-to-u} we automatically have $\det_q ({M})= \1$ (\cite{ps}).

\subsection{$\mathbf{L}^{\pm}$-operators}

Below we explicitly list the entries of the $\mathbf{L}^{\pm}$-operators 
obtained from the standard universal $\cR$-matrices of $\qAN$ by means 
of the algorithm of Faddeev, Reshetikhin and Takhtajan (\cite{noumi}):
\bgt
\forall_{i,j \in \overline{1,N+1}} \ : \ \left\{ \barr{ll}
L^{+}_{i i} = \kp{i} = \lb L^{-}_{i i} \rb^{-1}, & \\ \\
L^{+}_{i j} = 0 = L^{-}_{j i} & \rm{for} \ i > j, \\ \\
L^{+}_{i j} = \la \kp{i} E_{j i} & \rm{for} \ i < j, \\ \\
L^{-}_{i j} = - \la E_{j i} \km{j} & \rm{for} \ i > j,
\earr \right. \label{Lop}
\end{gather}
where we have introduced the recursively defined symbols:
\bgt
\forall_{i,j \in \overline{1,N}} \ : \ \left\{ \barr{ll}
E_{i,i+1} = E_i, & \\ \\
E_{i,j+1} = E_{i j} E_j - q E_j E_{i j} & \rm{for} \ i < j, \\ \\
E_{i+1,i} = F_i, & \\ \\
E_{i+1,j} = E_{i+1,j+1} F_j - q^{-1} F_j E_{i+1,j+1} & \rm{for} \ i >
j.
\earr \right. \label{eij}
\end{gather}
The above $\mathbf{L}^{\pm}$-operators are known (\cite{noumi}) to
satisfy the following relations:
\bgt
\mathbf{R}^{+}_{12} \mathbf{L}^{\pm}_1 \mathbf{L}^{\pm}_2 =
\mathbf{L}^{\pm}_2 \mathbf{L}^{\pm}_1 \mathbf{R}^{+}_{12}, \\ \non
\mathbf{L}^{+}_1 \mathbf{R}^{+}_{12} S \mathbf{L}^{-}_2 = S
\mathbf{L}^{-}_2 \mathbf{R}^{+}_{12} \mathbf{L}^{+}_1,
\end{gather}
and hence also
\bgt
\mathbf{R}^{+}_{21} S \mathbf{L}^{\pm}_1 S \mathbf{L}^{\pm}_2 = S
\mathbf{L}^{\pm}_2 S \mathbf{L}^{\pm}_1 \mathbf{R}^{+}_{21}, \\ \non
\mathbf{L}^{+}_2 \mathbf{R}^{+}_{21} S \mathbf{L}^{-}_1 = S
\mathbf{L}^{-}_1 \mathbf{R}^{+}_{21} \mathbf{L}^{+}_2,
\end{gather}
with
\beq
\R^{+} = \sum_{1 \leq i,j \leq N+1} q^{\d_{ij}} \me_{ii} \ox \me_{jj} 
+ \la \sum_{1 \leq i < j \leq N+1} \me_{ij} \ox \me_{ji}. \label{Rplus}
\end{equation}
At this stage it is a matter of an elementary algebra to verify that
the fundamental $\M$-matrix, \eqref{re-to-u}, with the operator entries:
\beq
M_{i j} = \sum_{k = 1}^{N+1} L^{+}_{i k} S L^{-}_{k j}, \label{MAN}
\end{equation}
does indeed satisfy \eqref{REprim} with $\R = \R^{+}$.

Finally, upon substituting \eqref{Lop} in the above formula and rearranging 
the resulting expressions we derive
\bgt
M_{i j} \big\vert_{i>j} = (-1)^{j+1-i} \la k_{2\ep_i} \left[
\tilde{E}_{j i} + q^{-2} \la \sum_{i < k \leq N+1} (-1)^{i-k} \km{i}
\kp{k} E_{k i} \tilde{E}_{j k} \right], \label{gMAN} \\ \non
M_{i j} \big\vert_{i<j} = q^{-1} \la \kp{i} \kp{j} \left[ E_{j i} +
\la \sum_{j < k \leq N+1} (-1)^{k+1-j} \km{j} \kp{k} E_{k i}
\tilde{E}_{j k} \right], \label{lMAN} \\ \non
M_{i i} = k_{2 \ep_i} \left[ 1 + q^{-1} \la^2 \sum_{i < k \leq N+1}
(-1)^{k+1-i} \km{i} \kp{k} E_{k i} \tilde{E}_{i k} \right],
\label{DMAN}
\end{gather}
with the operators $\tilde{E}_{i j}$ defined in analogy to \eqref{eij},
\bgt
\forall_{i,j \in \ovl{1,N}} \ : \ \left\{ \barr{ll}
\tilde{E}_{i,i+1} = E_i, & \\ \\
\tilde{E}_{i,j+1} = q^{-1} E_j \tilde{E}_{i j} - \tilde{E}_{i j} E_j &
\rm{for} \ i < j,
\earr \right. \label{tildeij}
\end{gather}
and further related to the $E_{i j}$'s through
\beq
\forall_{i,j \in \ovl{1,N+1}, \; i < j} \ : \ S E_{i j} =
(-1)^{j-i} \km{i} \kp{j} \tilde{E}_{i j}. \label{seij}
\end{equation}
Amongst the $(N+1)^2$ entries of the $\mathbf{M}$-matrix there is a
distinguished group of
the diagonal ones, $M_{i i}$, of which $N$ can be chosen to commute
with one another\footnote{The remaining one is the Casimir operator 
$\gt{c}_1$.} and therefore span the Cartan subalgebra of $\reAN$.
We shall have need for them in the sequel.

The convention on the $\mathbf{L}^{\pm}$-operators just
displayed proves exceptionally convenient for the analysis to
come. Finally, let us also note that the Casimir operators of the underlying
$\reAN$ translate naturally into Casimir
operators of $\qANex$ described by means of
\eqref{gMAN}-\eqref{DMAN},
\beq
\mathfrak{c}_k \to \cC_k. \label{Cas-to-Cas}
\end{equation}

\subsection{Representations of $\reAN$.}\label{app-inequiv}

The irreducible \hw \reps $\cR^L_\La$ of $\reAN$ induced by \eqref{re-uqa} 
are of the form: 
\beq\label{rep-ex-in}
\Rep_{\tx{ind.}} ( \reAN ) = \bigoplus_{\La \in \faffN} \bigoplus_{L \in 
\bZ_{N+1}} R^L_{\La}, \quad R^L_{\La} \sim R^{l , \om}_{\La}\vert_{L(l,\om) 
= L \mod 2}.
\end{equation}
Below we demonstrate that the \repsn: $R^L_{\La}, \; ( \La , L ) \in \faffN 
\x \bZ_{N+1}$ are indeed pairwise inequivalent.

By \eqref{out-ck} and \eqref{aff-rot-sc}
all $R^l_{\lb\om^{*}_N \rb^l \La}$ for $l\in\bZ_{N+1}$ have equal scalar
operators. It is therefore the latter that we focus on in the 
sequel\footnote{ The more general equivalence: $R^{l_1}_{\lb 
\om^{*}_N \rb^{l_1} \La_1} \sim R^{l_2}_{\lb \om^{*}_N \rb^{l_2} \La_2}$ 
is ruled out upon noting that it would - in particular - require that the 
Casimir eigenvalues for $R^0_{\La_1}$ and $R^0_{\La_2}$ be equal, which is 
obviously not the case unless $\La_1 = \La_2$.}, further reducing our 
problem by making the following observation:
\beq
\lb \exists_{\La \in \faffN} \exists_{l_1,l_2 \in \bZ_{N+1}} : R^{l_1}_{\lb
\om^{*}_N \rb^{l_1} \La} \sim R^{l_2}_{\lb \om^{*}_N \rb^{l_2} \La} \rb
\Longrightarrow R^{l_2 - l_1}_{\lb \om^{*}_N \rb^{l_2} \La} \sim R^0_{\lb
\om^{*}_N \rb^{l_1} \La},
\end{equation}
manifestly true in view of, e.g., the invertibility of the elementary \aut
$\om_{q,N}$. Accordingly, we next compare the eigenvalues of
\beq
M_{N+1,N+1} = k_{2 \ep_{N+1}}
\end{equation}
on $R^0_{\La}$ and $R^l_{\lb\om^{*}_N \rb^l \La},\ l \neq 0$.
To these ends we take a general state of the module $\cH_{\La}$ and -
respectively - of the module $\cH_{\lb \om^{*}_N \rb^l \La}$
($V_0, V_l$ are the corresponding \hw states),
\beq
\cH_{\La}\ni\v_{(m^0)}
\underset{\tx{\ciut{symb.}}}{\sim} F_1^{m^0_1} \ldots
F_N^{m^0_N} \act V_0 \quad ,\quad
\cH_{\lb \om^{*}_N \rb^l \La} \ni \v_{(m^l)}
\underset{\tx{\ciut{symb.}}}{\sim} F_1^{m^l_1} \ldots
F_N^{m^l_N} \act V_l,
\end{equation}
with all $m^0_j, m^l_j \leq \sum_{i=1}^N \la_i$. Using \eqref{espc-a} we 
then verify that
\beq
M_{N+1,N+1} \act \v_{(m^n)} = q^{h(n)} \v_{(m^n)}, \quad n \in \{ 0, l \}
\end{equation}
for
\beq
h(n):=2 m^n_N - \frac{2 \k_N n}{N+1} - \frac{2s_n(\La)}{N+1},
\end{equation}
in which
\beqa
s_0(\La) &:=& \sum_{k = 1}^N k \la_k,\non \\
s_l(\La) &:=& \sum_{k = 1}^N k
\om_N^l(\la_k)=s_0(\La) - (N+1) \sum_{k = l}^N \la_k + ( N + 1 - l ) \k.
\eeqa
In the present notation the condition of equivalence of the two \irreps boils 
down (in view of $m^L_N\leq\sum_{i=1}^N\la_i$) to the following statement:
\beq\label{xx}
\forall_{0 \leq m^0_N \leq s_0'(\La)} \ \exists_{0 \leq m^l_N \leq
s_l'(\La)} \ : \ m^l_N - m^0_N + \frac{s_0(\La) - s_l(\La)}{N+1} - \frac{l
\k_N}{N+1}=0 \mod \k_N,
\end{equation}
where
\beq
s_0'(\La) = \sum_{k = 1}^N \la_{k} \quad , \quad s_l'(\La) := \sum_{k = 1}^N
\lb \a_k , (\om^*_N)^l \La \rb = \k - \la_l.
\end{equation}
An easy computation then shows the left hand side of \eqref{xx} to be equal 
to
\beq\label{xxx}
m^l_N - m^0_N + \sum_{k = l}^N \la_k - l - \k
\end{equation}
and so, choosing\footnote{The existence of a nontrivial state for any such     
choice of $m^0_N$ follows from the Weyl symmetry of irreducible 
$\qU2$-submodules of the given $\reAN$-module.} $m^0_N := \sum_{k = l}^N
\la_k \leq s_0'(\La)$ for an arbitrary $l \in \overline{1,N}$ and using 
$m^l_N \leq s_l'(\La) \leq \k$, we arrive at the inequality:
\beq
-\k - l \leq \eqref{xxx} < 0.
\end{equation}
Thus \eqref{xx} is showed to be false which completes our proof of
mutual inequivalence of the \irreps \eqref{rep-ex-in}. $\square$

\end{document}